\documentclass[reprint, superscriptaddress, aps, pre]{revtex4-2}

\usepackage{graphicx}
\usepackage{bm}

\usepackage{color}
\usepackage{amsmath}
\usepackage{amssymb}
\usepackage[T1]{fontenc}

\def\vec#1{{\bm{#1}}}
\def\mat#1{{\hat{\vec{#1}}}}
\def\t#1{{\mathrm{#1}}}
\def\H{{\mathcal{H}}}
\def\vp{{\varphi}}
\def\e{{\varepsilon}}

\begin{document}


\title{Three-magnon scattering of spin wave on edge-localized mode in thin ferromagnetic film}

\author{Julia Kharlan}
\email{yulkha@amu.edu.pl}
\affiliation{ISQI, Faculty of Physics and Astronomy, Adam Mickiewicz University, Poznań, Poland}%
\affiliation{V. G. Baryakhtar Institute of Magnetism of the NAS of Ukraine, Kyiv, Ukraine}%

\author{Roman Verba}
\affiliation{V. G. Baryakhtar Institute of Magnetism of the NAS of Ukraine, Kyiv, Ukraine}%

\author{Krzysztof Sobucki}
\email{krzsob@amu.edu.pl}
\affiliation{ISQI, Faculty of Physics and Astronomy, Adam Mickiewicz University, Poznań, Poland}%

\author{Pawe\l{} Gruszecki}
\affiliation{ISQI, Faculty of Physics and Astronomy, Adam Mickiewicz University, Poznań, Poland}%

\author{Maciej Krawczyk}
\email{krawczyk@amu.edu.pl}
\affiliation{ISQI, Faculty of Physics and Astronomy, Adam Mickiewicz University, Poznań, Poland}%

\date{\today}

\begin{abstract}
 Three-wave scattering is a fascinating phenomenon with many applications in various technologies. Reducing the system symmetry greatly affects three-wave scattering, which, in this case, goes beyond the simple momentum conservation law. In this study, we examine three-magnon scattering at the edge of a thin ferromagnetic film, when a bulk spin wave interacts with an edge-localized propagating spin-wave upon the reflection. This creates new bulk spin waves at mixed frequencies by means of three-magnon confluence or stimulated splitting processes. Using our developed analytical theory, which has been confirmed by full micromagnetic simulations, we demonstrate that the amplitude of the wave generated in the stimulated splitting process is several times larger than that generated in the confluence process, primarily due to the lower group velocity. Furthermore,  intensity of inelastically scattered waves exhibit a pronounced dependence on the incidence angle and frequency of the edge spin wave that goes beyond existing qualitative models. We show that the observed behaviors can only be explained by taking into account, that the scattered waves are created by several elementary three-magnon  processes involving the incident and reflected waves. The complex nature of the scattered wave creation results in a strong sensitivity of its amplitude to the phase accumulation of spin waves upon reflection. 
\end{abstract}

\maketitle


\section{\label{sec:intro} Introduction}

Spin waves (SWs), excitations of magnetization order, are known to exhibit a variety of nonlinear processes, including  parametric interaction with electromagnetic pumping, three-wave, four-wave and higher order magnon-magnon interaction processes~\cite{Gurevich_Book1996, Lvov_Book1994, Zheng_JAP2023}. These nonlinear SW interactions manifest themselves in various phenomena, such as parametric SW instability~\cite{Suhl_JPCS1957, Patton_PSS1979}, saturation of the ferromagnetic resonance and foldover effect~\cite{Suhl_JAP1960, Gottlieb_JAP1962}, nonlinear decay of SWs~\cite{Boardman_PRB1988, Dobin_PRL2003}, and SW turbulence and chaos~\cite{Gibson_PRA1984, Aguiar_PRL1986, Lvov_Book1994}. Many nonlinear interactions become pronounced at relatively low and moderate SW power, which constitutes a major advantage of SWs in magnetically-ordered materials as opposed to phonons, photons, polarons, and many other excitations in microwave frequency band. Exploration of nonlinear SW phenomena enabled the development of many nonlinear microwave signal processing devices, such as frequency-selective power limiters, signal-to-noise enhancers, frequency mixers, correlators, and others~\cite{Stitzer_CSSP1985, Ishak_IEEEProc1988, How_Book2005}. The miniaturization of nonlinear magnetic microwave devices results in considerable investigations of nonlinear SW interactions at micro- and nanoscale, which was found to differ from that in bulk samples and thick films~\cite{Zheng_JAP2023, Verba_Chapter2024}, studied decades ago. 

Recently, interest of nonlinear SW dynamics at micro- and nanoscale has been additionally stimulated by a rapidly grown field of magnon computing~\cite{Chumak_NatPhys2015, Pirro_NRM2021, Chumak_Roadmap2022}, which promises energy-efficient logic and novel wave-based beyond-Boolean computing elements and circuits. It is worth noting that nonlinearity is crucial for the development of both the logic elements~\cite{Wang_NE2020, Wang_PRAppl2024} and true wave-based neural networks~\cite{Papp_NC2021}, as using only SW interference, diffraction and other linear phenomena does not allow for all-magnon realisation of the most from these functionalities. 

The three-magnon interaction processes are the lowest-order multi-magnon processes, usually the first ones, which  become pronounced with an increase in SW power (if these processes are allowed). A familiar example of three-magnon scattering is the first-order Suhl instability~\cite{Suhl_JPCS1957}, which is a splitting of a quasi-uniform magnon (a.k.a. ferromagnetic resonance, FMR, mode) into two counter-propagating magnons if the amplitude of the first exceeds a certain threshold. Suhl instability, if allowed by the spectrum, is often the primary mechanism which limits maximal amplitude of FMR at high powers. Similar process of three-wave splitting of a primary propagating SW, called ``three-wave decay instability''~\cite{Boardman_PRB1988}, often limits maximal power, which can be transmitted by SWs. This process forms the basis for microwave frequency-selective limiters. 

The converse process is the three-magnon confluence, when two magnon of the same or different kinds fuse to form a single resultant magnon. This process manifests itself as the generation of second harmonics~\cite{koerner2022, kumar2024, Demidov_PRB2011} or frequency mixing, as well as is responsible for resonant nonlinear damping and inversion of a nanomagnet response to a spin torque~\cite{Barsukov_SciAdv2019}. In contrast to spontaneous splitting, the confluence is a thresholdless process that occurs at any amplitude of the primary magnon pair. 

Three-wave splitting processes can be spontaneous (when only the primary magnon is excited), but also  stimulated, when one of the secondary (split) magnons is excited by a separate external stimulus. For example, the splitting of a magnon 1 into magnons 2 and 3 (in short, $1\to 2+3$), which satisfies the energy conservation rule $\omega_1 = \omega_2 + \omega_3$, will be a stimulated process if the magnon 2 (for definiteness, the same holds for the magnon 3) is also externally excited in addition to the magnon 1. The stimulated process is thresholdless, and results in the appearance of the magnon 3 and increase of the number of the magnon 2, i.e., amplification of the magnon 2, which is interesting for magnon transistor applications~\cite{ge2024}. 


The three-magnon interaction between plane SWs in bulk samples and ferromagnetic films has been thoroughly studied in the past century~\cite{Lvov_Book1994, Wigen_Book1994}. Many specific features of the three-wave interaction in thin films have also been revealed~\cite{ordonez2009, liu2019, qu2023}. Recent research has uncovered distinct features of three-magnon processes in magnetic nanostructures, in particular, specific selection rules~\cite{Schultheiss_PRL2019, Camley_PRB2014} and the ability for efficient, mode-selective control of these processes~\cite{Etesamirad_PRAppl2023, Verba_Chapter2024}, as well as the potential to implement the stimulated splitting~\cite{Korber_PRL2020} for neuromorphic computing~\cite{Korber_NC2023, heins2025}. The stimulated three-magnon interaction between plane SWs and magnetic solitons -- domain wall, vortex, skyrmion -- has mostly been studied in relation to the development of frequency combs~\cite{Yao2023, wang2021, zhou2021, wang2022}.

In the previous works~\cite{Pawel2022, sobucki2025goos}, we demonstrated three-magnon scattering of a bulk SW beam on an edge-localized SW in a thin ferromagnetic film. This process, resulting in the appearance of another scattered bulk SW via confluence or splitting processes, is interesting for various magnonic applications, for instance, frequency demultiplexing. Using micromagnetic simulations, we identified several significant characteristics of three-wave scattering involving both bulk and edge SWs, such as its high angular sensitivity and the pronounced asymmetry in the efficiency of confluence and splitting processes. However, the proposed analytical model could not explain these observations.

In this paper, we develop a detailed theory of the three-magnon scattering of bulk and edge SWs in a thin ferromagnetic film. In addition to the successful explanation of previous observations, the theory uncovers that the inelastic scattering of an SW beam on an edge mode is more complex than a simple single-stage process. It also predicts a high sensitivity of the nonlinear scattering to the phase relation between inelastically scattered SWs before and after reflection from the film edge. This feature could result in an effective vanishing (in fact, compensation) of the three-wave scattering, which is confirmed in our micromagnetic simulations and constitutes an additional method of controlling the scattering. 

The paper is organized as follows. In Sec.~\ref{Sec:Structure} we describe the system under study. In the next section, we present a theory of three-magnon scattering of bulk and edge SWs, starting with the basics of the vectorial Hamiltonian formalism for nonlinear SW dynamics (Sec.~\ref{ss:basic-eq}), analyzing a generic case of three-magnon interaction of two bulk SWs and an edge mode (Sec.~\ref{ss:V123-calc}), and finishing with the calculation of scattered wave amplitudes, which were found to be sensitive to SW group velocity and wave interference (Sec.~\ref{ss:phi-ac}). Section~\ref{s:res} presents results of micromagnetic simulations, their comparison with theory and related analysis. 

\section{System under study}\label{Sec:Structure}

\begin{figure}
	\includegraphics[width=\columnwidth]{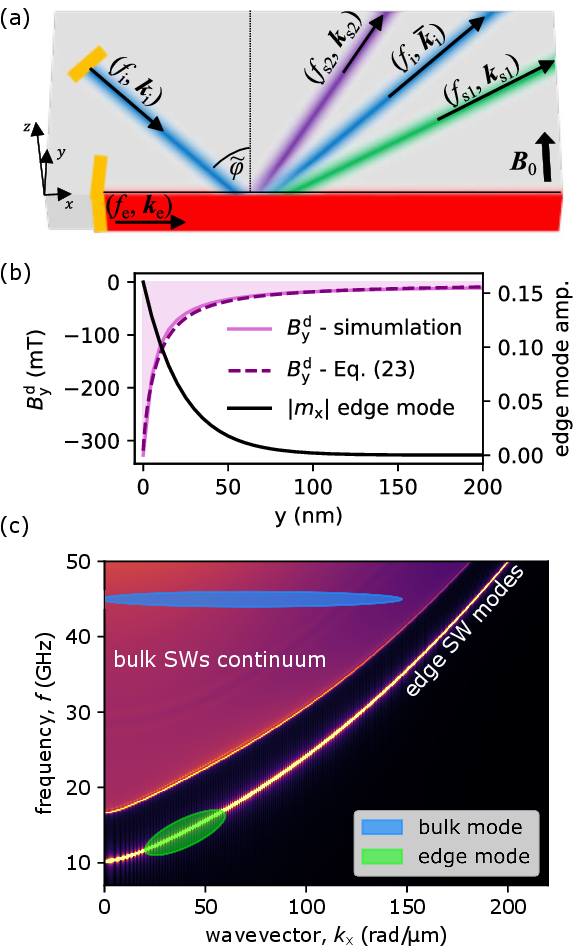}
	\caption{(a) A sketch of the system under study. A semi-infinite thin Py layer ($h=10$~nm) is subjected to a uniform external magnetic field $B_0=300$~mT, parallel to the $y$~axis. Two antennas excite incident bulk SW with the frequency $f_\t{i}$ ($f_\t{i}=\omega_\t{i}/2\pi$) and wave vector $\vec k_\t{i}$ (blue beam) and edge SW at the frequency $f_\t{e}$ having wave vector $\vec k_\t{e}$. Standard reflection of the bulk SW, incident at the angle $\tilde \vp$ (beam incidence angle is shown), results in the appearance of the reflected beam at $(f_\t{i}, \bar{\vec k}_\t{i})$, while three-magnon interaction of incident bulk SW with the edge mode -- confluence and stimulated splitting -- creates two inelastically scattered waves $(f_{\t{s}1},\vec k_{\t{s}1})$ (green) and $(f_{\t{s}2},\vec k_{\t{s}2})$ (purple); mutual position of reflected, fused and split beams may vary with the incidence angle and SWs frequencies. 
     (b) Demagnetizing field distribution along the $y$~axis calculated with numerical simulation (the purple field) and derived with an analytical formula Eq.~(\ref{eq:demag_field_formula}) (dark violet dashed line). The amplitude of edge SW magnetization $x$-component ($|m_x|$) along $y$~axis at 12 GHz is shown with the solid black like.
    (c) The dispersion relation of the SWs along the $k_x$ axis, which may be divided into two parts.  The blue and green colours indicate the range of bulk SW wavenumbers (i.e., $0 \leq \varphi \leq 55^\circ$) and the frequency/wavenumber range of edge SWs considered in the paper.  } 
	\label{fig:geo}
\end{figure}

The system under investigation consists of a permalloy (Py, $\mathrm{Ni}_{80}\mathrm{Fe}_{20}$, the saturation magnetization $M_s=800$~kA/m, the exchange stiffness constant $A=13$~pJ/m and the gyromagnetic ratio $\gamma=176\cdot10^9$~rad/sT) semi-infinite thin film (thickness $h=10$~nm), magnetized by an in-plane bias magnetic field $\vec B_0 =  B_0 \vec e_y$ ($B_0=0.3$~T), applied perpendicularly to the film edge, as shown in  Fig.~\ref{fig:geo}a. In this geometry, the nonuniformity of the demagnetization field at the film edge creates a field well, where the low frequency edge SWs are confined~\cite{Chia_PRB2012}. Figure~\ref{fig:geo}{b} shows the profiles of the demagnetising field and the profile of the edge SW (at 12 GHz) along the $y$ axis, respectively.

We study the bulk SWs incident on the film edge, along which the edge SW propagate, and the inelastic scattering of the bulk SW on the edge SW (see the schematic plot in Fig.~\ref{fig:geo}a). We assume that both bulk and edge SWs are excited independently by two antennas, at angular frequencies $\omega_\t{i}$ and $\omega_\t{e}$, respectively, where $\omega_\t{i} >  \omega_\t{e}$. Propagation direction (incidence angle) of the bulk SW is determined by the orientation of antenna, which defines the phase velocity angle $\vp = \t{arctan}[k_{\t{i},x}/(-k_{\t{i},y})]$ ($k_{\t{i},x}$ and $k_{\t{i},y}$ are the components of the bulk SW). SW wave number $k_\t{i} = |\vec k_\t{i}|$ is defined by the excitation frequency and the angle via the dispersion relation. The same applies to edge mode as well, with the difference that its wave vector is one-dimensional, $\vec k_\t{e} = k_\t{e} \vec e_x$.  Anisotropy of SW dispersion in an in-plane magnetized film leads to a certain difference between the direction of phase and group velocities of bulk SWs, thus, between the beam incidence angle $\tilde\vp$ (shown in Fig.~\ref{fig:geo}a) and phase-defined angle $\vp$. Below, we always use only the phase-defined angle $\vp$.

The three-magnon interaction between the bulk and edge SWs results in the appearance of inelastically scattered SWs: the confluence process creates a fused SW at the frequency $\omega_\t{s} = \omega_\t{i} + \omega_\t{e}$, while the splitting process is evidenced by the presence of split SW at $\omega_\t{s} = \omega_\t{i} - \omega_\t{e}$ (we study the case $\omega_\t{i} >  \omega_\t{e}$).
Hereinafter, the index ``s'' stands for a scattered wave, both for fused and split SWs.  The wave vectors of the scattered SWs are defined by the momentum conservation law and SW dispersion, as described below. Naturally, an elastically reflected SW also appears at the same frequency $\omega_\t{i}$ as the incident bulk SW and wave vector $\bar{\vec k}_\t{i} = (k_{\t{i},x},-k_{\t{i},y})$.

\section{Theoretical description}\label{s:th}

\subsection{Principal equations}\label{ss:basic-eq}

For a theoretical description of the three-magnon scattering of bulk and edge SWs we use the recently developed vectorial Hamiltonian formalism~\cite{Tyberkevych_ArXiv}. In comparison with well-known scalar Hamiltonian formalism for SW dynamics, which is based on the classical analogue of Holstein-Primakoff transformations~\cite{Holstein_PR1940}, the vectorial formalism  allows for convenient treatment of nonuniform static magnetization distributions as well as of complex (non-plane-wave) SW profiles. The latter feature is of a particular importance for our study, since the edge SW mode is not a plane wave in all dimensions (see its profile in Fig.~\ref{fig:geo}b).

Within vectorial Hamiltonian formalism, time- and space-dependent three-dimensional magnetization vector $\vec M(\vec r,t)$ is represented as
	\begin{equation}\label{e:VHF-map}
		\frac{\vec M(\vec r,t)}{M_s} = \left(1-\frac{|\vec{s}(\vec{r},t)|}{2}\right)\vec{\mu}(\vec{r}) + \sqrt{1-\frac{|\vec{s}(\vec{r},t)|}{4}}\vec{s}(\vec{r},t) \,,
	\end{equation}
where $\vec\mu(\vec r) = \vec M(\vec r)/M_s$ is the spatial distribution of the normalized static magnetization, and $\vec s(\vec r,t)$ is the dimensionless dynamic magnetization, which is perpendicular to the static magnetization in each space point, $\vec s \perp \vec\mu$. The dynamic magnetization can be expanded into a series of linear SW eigenmodes $\vec s_\nu(\vec r)$, which can be either standing modes or propagating waves, depending on the system: 
	\begin{equation}\label{e:s-expansion}
		\vec{s}(\vec{r},t)=\sum_{\nu}[c_\nu(t)\vec{s}_\nu(\vec{r})+ \t{c.c.}], 
	\end{equation}
where $c_\nu(t)$ are the time-dependent complex amplitudes of the SW modes. Linear SW modes possess the orthogonality relation 
	\begin{equation}\label{e:norm}
		\frac{1}{V} \int \vec{s}_{\nu'}^{*}\cdot\vec{\mu}\times\vec{s}_{\nu} d\vec{r} = -i A_\nu \Delta_{\nu,\nu'}, 
	\end{equation}
where the integration is performed over the sample volume $V$ and $A_\nu$ is the normalization constant and $\Delta_{ij}$ is the Kronecker delta. While it is often convenient to normalize SW mode profiles $\vec s_\nu$ to get $A_\nu = 1$, it is not the case of our study because bulk and edge SWs have different dimensionality. 

The next step is to represent the Hamiltonian function of the system in terms of the SW amplitudes $c_\nu(t)$. In ferromagnetic systems, it is convenient to use the normalized Hamiltonian $\H=\gamma E/(M_sV)$, where $E$ is the total magnetic energy. This ensures that the Hamiltonian function and all its expansion coefficients have the same units of frequency~\cite{Krivosik_PRB2010}. Dynamics of SW amplitudes is then given by Hamiltonian equations~\cite{Tyberkevych_ArXiv}:
	\begin{equation}\label{e:dc-dH}
		\frac{dc_\nu(t)}{dt} = -\frac{i}{A_\nu} \frac{\partial \H}{\partial c_\nu^*(t)} .
	\end{equation}
In our study, we are interested in three-magnon confluence and splitting, and thus the relevant component in SW Hamiltonian reads:
	\begin{equation}\label{e:H3-gen}
		\H^{(3)} = \sum_{123}(V_{12,3} c_1c_2c_3^*+\t{c.c.}).
	\end{equation}
The term $V_{12,3}c_1c_2c_3^*$ describes confluence process of the magnons 1 and 2 into the magnon 3, while the complex conjugated term describes the reverse splitting process $3\to 1+2$. Three-magnon coefficient is given by~\cite{Tyberkevych_ArXiv, Verba_Chapter2024}:
\begin{equation}\label{e:V123}
	\begin{split}
		V_{12,3}&=-\frac{\omega_M}{2V}\int \left[ (\vec s_1 \cdot \vec s_2) \vec\mu \cdot \mat N \cdot \vec s_3^* \right. \\
		&+ \left. (\vec s_1 \cdot \vec s_3^*) \vec\mu \cdot \mat N \cdot \vec s_2  + (\vec s_2 \cdot \vec s_3^*) \vec\mu \cdot \mat N \cdot \vec s_1 \right] d\vec r . 
	\end{split}
\end{equation}
Here, $\omega_M=\gamma\mu_0 M_s$ and $\mat N$ is a tensor operator describing magnetic interactions, which is a sum of different contributions, such as exchange, dipolar, anisotropy, etc. In the case of uniform static magnetization, the exchange interaction does not contribute to the three-magnon interaction~\cite{Verba_Chapter2024}. The negligible magnetocrystalline anisotropy in Py results in the dipolar interactions being the only relevant contributions:
	\begin{equation}\label{e:N-dip}
		\mat N(\vec r) \cdot \vec s_\nu (\vec r) \equiv
        \mat N^\t{dip}(\vec r) \cdot \vec s_\nu (\vec r) = \int \mat G(\vec r,\vec r') \cdot \vec s_\nu (\vec r') d \vec r' ,
	\end{equation}
where $\mat G$ is the magnetostatic Green function~\cite{Guslienko_JMMM2011}.

\subsection{Efficiency of three-magnon interaction of bulk and edge waves}\label{ss:V123-calc}

Following a generic approach, described in Sec.~\ref{ss:basic-eq}, we first calculate three-magnon interaction coefficients. In this subsection, we derive the formula for the coefficient $V_{12,3}$, where SWs 1 and 3 are arbitrary bulk waves and 2 is an edge-localized SW. Such a general case includes both (bulk$+$edge$\to$bulk) confluence and (bulk$\to$bulk$+$edge) splitting (including stimulated splitting, i.e., when the edge SW is present during the scattering); other possible alternatives (e.g., interaction of two edge modes with one bulk or splitting of edge mode into two bulk ones) can be investigated in the same manner, but expression, derived below, are not applicable to them directly.

We assume that the bulk SWs are 2D plane waves, which profiles in Cartesian components $\vec s_j = (s_{j,x},s_{j,y},s_{j,z})$ can be parametrized as follows:
	\begin{equation}\label{e:s-bulk}
		\vec s_j = (1,0,i \e_j) \exp\left[i(k_{j,x}x + k_{j,y}y) \right]
	\end{equation}
where $j = 1,3$ is the index denoting bulk SW, coefficient $\e_j \in \mathbb{R}$ describes magnetization precession ellipticity. The profile across the film thickness is assumed to be uniform as it holds for fundamental mode of thin films. The profile in Eq.~(\ref{e:s-bulk}) is not normalized, but it is not needed as was  pointed out above. Ellipticity coefficient as well as the dispersion relation $\omega_{\vec k_j}$ of bulk SWs in uniformly magnetized thin ferromagnetic films are well-known~\cite{Kalinikos_JPC1986}:
	\begin{equation}\label{e:def-omega-e}
		\omega_{\vec k_j} = \sqrt{\Omega_\t{IP} \Omega_{zz}}\,, \quad \e_j = \sqrt{\frac{\Omega_\t{IP}}{\Omega_{zz}}} ,
	\end{equation}
where 
	\begin{subequations}
		\begin{equation}
			\Omega_\t{IP} = \omega_H + \omega_M \lambda^2 k_j^2 + \omega_M \tilde f(k_jh) \sin^2\vp_j \,,
		\end{equation}
		\begin{equation}
			\Omega_{zz} = \omega_H + \omega_M \lambda^2 k_j^2 + \omega_M \left(1-\tilde f(k_jh) \right) \,,
		\end{equation}
	\end{subequations}	
where $\omega_H = \gamma B_0$, $\lambda=\sqrt{2A(\mu_0 M_s^2)^{-1}}$ is the exchange length, $k_j \equiv |{\vec k}_j|$, $\tilde f(x)=1-(1-e^{-|x|})/|x|$ is the so-called ``thin film function'' and $\vp_j = \t{arctan}[k_{j,x}/(-k_{j,y})]$ is the wave propagation angle, defined by the phase velocity as explained earlier. 

For the edge SW, we assume the profile as
	\begin{equation}\label{e:s-edge}
		\vec s_2= (1,0,i\e_2) e^{ik_{2,x}x} e^{-\kappa y} .	
	\end{equation}
The exponential law $e^{-\kappa y}$ is an approximation, however, it reasonably describes the real micromagnetic edge mode profile (see the inset in Fig.~\ref{fig:geo}b) and enables a more in-depth analytical treatment of the three-magnon interaction. The inverse localization length, $\kappa$, and the precession ellipticity,       $\e_2$, as functions of SW frequency, will be taken from micromagnetic simulations. 

Recalling that the static magnetization in the considered case is $\vec\mu (\vec r) = \vec e_y$, one finds that only off-diagonal $yx$-component of the magnetic interaction tensor $\mat N$ contributes to the three-magnon coefficients, Eq.~(\ref{e:V123}). In thin-film case (uniform magnetization across the thickness) the magnetostatic Green function in Eq.~(\ref{e:N-dip}) can be conveniently represented as
	\begin{equation}
		\mat G(\vec r, \vec r') = \frac{1}{4\pi^2} \int \mat N_\vec q e^{i\vec q\cdot (\vec r - \vec r')} d^2 \vec q 
	\end{equation}   
with the relevant component of the Green function in the wave-vector, $\vec q$, space~\cite{Guslienko_JMMM2011}
	\begin{equation} \label{eq:N_dip}
		N_{\vec q}^{yx} = \frac{q_x q_y}{q^2} \tilde f(qh).
	\end{equation}
As we turn to the effectively 2D case, the $\vec r$ and $\vec q$ are two-dimensional vectors of the position and momentum space. Consequently, the integration over sample volume $V$ should be changed to integration over the area $S$ in Eqs.~(\ref{e:norm}), (\ref{e:V123}), and (\ref{e:N-dip}).  

Let's inspect in details the first term in the expression (\ref{e:V123}) for the thee-magnon coefficient, $V_{12,3}$. By substituting the generic profiles of SWs, Eqs.~(\ref{e:s-bulk}) and Eqs.~(\ref{e:s-edge}), it can be represented as follows:
	\begin{equation}
		T_1 = - \frac{\omega_M}{2L_xL_y} (1-\e_1 \e_2) \int dq_x dq_y N_{\vec q}^{yx} I_x I_y \,,  
	\end{equation}
where
	\begin{equation}\label{e:Ix}
		I_x = \int\limits_{-\infty}^\infty dx \int\limits_{-\infty}^\infty dx' e^{i(k_{1,x}+k_{2,x})x} e^{iq_x (x-x')} e^{-ik_{3,x}x'},
	\end{equation}
    and
	\begin{equation}\label{e:Iy}
		I_y = \int\limits_{0}^\infty dy \int\limits_{0}^\infty dy' e^{(ik_{1,y} - \kappa)y} e^{iq_y (y-y')} e^{-ik_{3,y}y'} .
	\end{equation}
Here, $L_x$ and $L_y$ are formal film sizes in the $x$ and $y$ directions, assumed to be sufficiently large $L_x,L_y \to \infty$. These quantities are present in the intermediate steps, but will be shortened in the final expressions. The first integral is trivial and yields $I_x = 2\pi L_x \delta(k_{1,x} + k_{2,x} - q_x) \Delta_{k_{1,x} + k_{2,x}, k_{3,x}}$, where the last Kronecker delta indicates the momentum conservation rule for the $x$ components of the SW wave vectors, given that the system possesses translational symmetry in the $x$ direction. The integration over $y$ in Eq.~(\ref{e:Iy})  is also trivial. However, this is not the case for the integration over $y'$. Here, we use the relation 
	\begin{equation}
		\int\limits_0^{\infty} e^{-i \xi y} dy = \lim_{\beta \to +0} \int\limits_0^{\infty} e^{-(i \xi + \beta)y} dy = -i \lim_{\beta \to +0} \frac{1}{\xi - i\beta}
	\end{equation}
followed by the application of the Sokhotski–Plemelj theorem~\cite{Plemelj_Book} for an arbitrary continuous function of the real variable $g(\xi)$:
	\begin{equation}
		\lim_{\beta \to +0} \int\limits_{-\infty}^\infty \frac{g(\xi)}{\xi - i\beta} = i\pi g(0) + \mathcal{P} \int\limits_{-\infty}^{\infty} \frac{g(\xi)}{\xi} d\xi \,, 
	\end{equation}
where $\mathcal{P}$ means an integral in the sense of the principal value. 

Using the above expressions, we calculate all the terms of three-magnon coefficient $V_{12,3}$, which can be represented as follows:
	\begin{equation}\label{e:V123-fin}
		\begin{split}
		V_{12,3} L_y = \tilde V_{12,3} = -\frac{\omega_M}{4}  \left[\left(1-\e_1\e_2\right) F(-\vec k_3, -1)  \right. \\
		+ \left(1+\e_1\e_3\right)  F(\vec k_2, +1) + \left. \left(1+\e_2\e_3\right) F(\vec k_1, -1) \right] \\
		\times \Delta_{k_{1,x} + k_{2,x}, k_{3,x}}.
		\end{split}
	\end{equation}
Here, we use the notation $\vec k_2 = k_{2,x} \vec e_x + k_{2,y} \vec e_y$, $k_{2,y} = k_{3,y} - k_{1,y}$, i.e., this is a wave vector component, which would be required by momentum conservation law if mode 2 was a bulk wave, and we define the function
	\begin{equation}\label{e:F-kp}
		F(\vec k, p) = \frac{N_{\vec k}^{yx}}{\kappa + i k_{2,y}} + \frac{ip}{\pi} \mathcal{P} \int\limits_{-\infty}^\infty \frac{N_{\vec k + q_y\vec e_y}^{yx}}{q_y \left(\kappa  + i(k_{2,y} + pq_y)\right)} dq_y.
	\end{equation} 

In Eq.~(\ref{e:V123-fin}) one finds the momentum conservation rule only for the $k_x$ component. The $y$ component of SW wave vector does not possess a conservation rule, because (i) the system has broken translational symmetry in the $y$ direction and (ii) the edge SW mode is characterized not by a single wave number $k_y$, but by a continuous spatial Fourier spectrum. This means that  in the three-magnon process the $k_x$ components of the interacting waves are determined by the momentum conservation rule, while the $k_y$ components are indirectly determined by the energy conservation rule. For instance, in the confluence process $(1+2)\to 3$ the energy conservation requires that $\omega_1 + \omega_2 = \omega_3$, and the $y$ component of the fused SWs wave vector $k_{3,y}$ is determined by $k_{3,x}$ and the dispersion relation $\omega_{{\vec k}_3}$. Thus, the confluence process is permitted over a broad spectrum of primary SWs (1 and 2) frequencies and incidence angles of the wave 1 -- the only requirement is that the lowest bulk SW frequency in the spectrum at a given $k_{3,x}$ is less than $\omega_1 + \omega_2$, $\min_{k_y} \omega(k_{1,x} + k_{2,x}, k_y) \lesssim (\omega_1 + \omega_2)$. A similar feature applies to the stimulated splitting process as well. This is in a sharp contrast with confluence and stimulated splitting three-magnon processes of bulk SWs, which are strongly resonant -- energy and momentum conservation rules in this case can be simultaneously satisfied only for a distinct set of frequencies and wave vectors of two primary SWs.

As it is evident from Eq.~(\ref{e:F-kp}), the interaction efficiency depends on the wave vectors of the interacting SWs. The largest three-magnon coefficient is expected when $k_{2,y} = 0$, i.e., when $k_{1,y} = k_{3,y}$, meaning that the wave vector components of the bulk SWs are close to ``momentum conservation rule'' when the edge SW is disregard. This is not something extraordinary, since the spatial spectrum of an exponentially localized edge SW has the largest intensity at $k_y = 0$. At the same time, wave number dependence of $|V_{12,3}|$ is expected to be weak if the difference in the wave numbers of the bulk SWs is less than inverse localization length of the edge one, $|k_{1,y} - k_{3,y}| < \kappa$, and, vice versa, it vanishes if $|k_{1,y} - k_{3,y}| \gg \kappa$.

\subsection{Amplitude of the scattered waves and the effect of phase accumulation}\label{ss:phi-ac}

\begin{figure}
\includegraphics[width=\columnwidth]{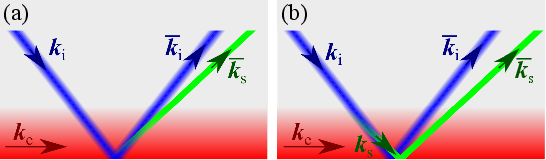}
    \caption{Schematic illustration of the two elementary scattering processes that contribute to the creation of inelastically scattered SWs: (a) three-magnon scattering of an elastically reflected SW with wave vector $\bar{\vec k}_\t{i}$ and an edge SW, leading to the formation of SW $\bar{\vec k}_\t{s}$, and (b) three-magnon scattering of an incident SW $\vec k_\t{i}$, which creates an inelastically scattered wave $\vec k_\t{s}$ with its subsequent reflection from the edge. The blue, red and green SW beams depict the incident (and linearly reflected), edge and inelastically scattered waves, respectively. } 
	\label{fig:2}
\end{figure}

Above, we derived the formula for the three-magnon coefficient for the interaction between two arbitrary bulk SWs and an edge wave. The creation of an inelastically scattered wave during the reflection of a bulk SW from an edge of the ferromagnetic film, where an edge mode is propagating, is a more complex process, which involves several elementary three-magnon processes. This is illustrated in Fig.~\ref{fig:2}. For definiteness, let's now consider the confluence. Before reflection from the film edge, an incident SW with the wave vector $\vec k_\t{i} = (k_{\t{i},x},-k_{\t{i},y})$ can fuse with an edge mode creating a co-propagating fused wave with $\vec k_\t{s} = (k_{\t{s},x}, -k_{\t{s},y})$, where $k_{\t{s},x}$ is determined by the momentum conservation law and $k_{\t{s},y}$ ($k_{\t{s},y}>0$) by the energy conservation law and the SW dispersion relation. This process is described by the coefficient $V_\t{ie,s}$, i.e., $V_{12,3}$ given in Eq.~(\ref{e:V123-fin}) with $\vec k_1 \equiv \vec k_\t{i}$, $\vec k_3 \equiv \vec k_\t{s}$, and mode 2 being the edge mode with the wave number $k_{\t{e},x}$. The resultant scattered  SW is then reflected from the edge and contributes to the inelastically scattered output signal. This process is shown in Fig.~\ref{fig:2}(b).

The second elementary process includes the incident wave linearly reflected from the film edge with a wave vector $\bar{\vec k}_\t{i} = (k_{\t{i},x}, +k_{\t{i},y})$, as shown  in Fig.~\ref{fig:2}(a). This reflected wave could also fuse with the edge SW (recall that the edge SW spans over certain length from the edge), resulting in the appearance of another co-propagating inelastically scattered wave having  $\bar{\vec k}_\t{s} = (k_{\t{s},x}, +k_{\t{s},y})$ (SW spectrum in our case is reciprocal, so SWs with wave vectors $\vec k_\t{s}$ and $\bar{\vec k}_\t{s}$ have the same frequency and, thus, both satisfy the energy conservation rule). This process, which occurs after the reflection of the incident SW, is described by the coefficient $V_{\bar{\t{i}}\t{e},\bar{\t{s}}}$. In the following, the bar over the symbols will be used to indicate bulk waves propagating in $+y$ direction. 

Two other elementary processes are also possible. As the three-magnon scattering of bulk and edge modes does not impose any direct restrictions on the $k_y$ components of the interacting waves, an incident wave with the wave vector $\vec k_\t{i}$ in the confluence process could create a ``counter-propagating'' (in the sense of the $k_y$ sign) fused wave at $\bar{\vec k}_\t{s}$. The same, linearly reflected SW at $\bar{\vec k}_\t{i}$ could result in the appearance of scattered wave at $\vec k_\t{s}$, which, upon reflection from the edge, also contributes to the total scattered wave at $\omega_\t{s} = \omega_\t{i} + \omega_\t{e}$. These two processes are described by the quantities $V_{\t{ie},\bar{\t{s}}}$ and $V_{\bar{\t{i}}\t{e,s}}$, respectively. Thus, in a general case, four elementary three-magnon scattering processes contributes to inelastic scattering of bulk SW from the edge. The last two mentioned elementary processes, however, are characterized by a large wave number difference $k_{2,y} \equiv k_{3,y} - k_{1,y} = \pm (k_{\t{i},y} + k_{\t{s},y})$ and are typically significantly weaker than the two earlier described contribution. For our numerical study (Sec.~\ref{s:res}) this also holds true, and below we neglect elementary processes generating ``counter-propagating'' waves.

The last effect we should take care of is the phase accumulation of the both incident and scattered bulk SWs. The first elementary process $(\t{i} + \t{e} \to \t{s})$ generates the scattered wave, which propagates towards to edge. Upon the reflection, it acquires a phase shift $\phi_\t{s}$; thus, as a result one observes a scattered wave proportional to $V_\t{ie,s} e^{i\phi_\t{s}}$. The second elementary process involves reflected incident SW, which is characterized by an additional phase shift $\phi_\t{i}$. Thus, the total ``effective'' three-magnon coefficient is not a simple sum of those describing elementary processes, but a phase-shift-weighted sum:
	 \begin{equation}\label{e:Vc-tot}
	 	V_\t{conf} = V_{\t{ie,s}} e^{i\phi_\t{s}} + V_{\bar{\t{i}}\t{e},\bar{\t{s}}} e^{i\phi_\t{i}}.
	 \end{equation}

Using the same procedure and recalling that for the stimulated splitting the conjugated three-magnon coefficient matter (see Eq.~\ref{e:dc-dH}), the total ``effective'' three-magnon efficiency is given by 
	 \begin{equation}\label{e:Vs-tot}
		V_\t{split}^* = V_\t{se,i}^* e^{i\phi_\t{s}} + V_{\bar{\t{s}}\t{e},\bar{\t{i}}}^* e^{i\phi_\t{i}}.
	\end{equation}
Here, the split wave has the frequency $\omega_\t{s} = \omega_\t{i} - \omega_\t{e}$, the wave number $k_{\t{s},x} = k_{\t{i},x} - k_{\t{e},x}$ and $k_{\t{s},y} < 0$ determined from the dispersion relation. The overbars in Eq.~(\ref{e:Vc-tot})-(\ref{e:Vs-tot}) stands, the same as before, for reflected split wave with $k_{\t{s},y}>0$.   
 
In the simplest case of homogeneous (in terms of magnetization and the magnetic field) thin ferromagnetic film and sharp edge, an SW acquires $\pi$ phase shift upon elastic reflection. In the studied geometry, however, SW behavior is more complex because of nonuniform static demagnetization fields near the film edge (see, Fig.~\ref{fig:geo}b), which can be  described by the expression~\cite{Aharoni}: 
	\begin{equation}\label{eq:demag_field_formula}
		B_y^\t{d}(y) = M_s \left(\frac{1}{2}-\frac1\pi \mathrm{arctan}\left[\frac{y}{h}\right]-\frac{y}{2\pi h} \log\left[1+\frac{h^2}{y^2}\right] \right) \ .  
	\end{equation}
This field inhomogeneity leads to the spatial dependence of the $y$ component of the SW wave vector, $k_{j,y}(y)$, which can be determined implicitly from the dispersion relation, Eq.~(\ref{e:def-omega-e}), at varying internal magnetic field, $\omega_{(k_{j,x},k_{j,y}(y))}(B_0 - B_y^\t{d}(y))$, at the frequency of incident or scattered SW, $\omega_\t{i}$ or $\omega_\t{s}$, respectively. The total phase accumulation is an integral of the wave number difference $(k_{j,y}(y) - k_{y,0}$ over SW propagation pass, where $k_{y,0} \equiv k_{j,y}(y\to \infty)$ means the  wave number far from the film edge. 

Rigorous definition of this quantity is, however, viably impossible. The problem is that it is impossible to define the exact position at which an elementary process of three-magnon scattering takes place, since both the localization length of the edge mode and the width of the field well are quantities of the same order and nature. We use the following estimation:
	\begin{equation}\label{e:phase}
		\phi \approx \pi + \int_0^{\infty}(k_{j,y}(y)-k_{y,0})dy,
	\end{equation}
which includes $\pi$ phase shift at the boundary (being the same for all the waves, and which does not affect the modulus of the effective three-magnon coefficients, see Eqs.~(\ref{e:Vc-tot}) and~(\ref{e:Vs-tot})). It also includes the additional phase accumulation of the SW acquired in the course of propagation from the boundary to the observation point (out of demagnetizing field nonuniformity). Although this is still quite a rough estimation, it is a natural average between two limiting cases: (i) assuming three-magnon scattering to occur exactly at the edge of the film (i.e., no additional phase accumulation), and (ii) assuming scattering to occur away from the field nonuniformity (it would correspond to multiplier of 2 before the integral in Eq.~(\ref{e:phase})), both of which are unrealistic limits. It is worth noting that the phase in Eq.~(\ref{e:phase}) depends on the SW frequency and thus is different for incident and inelastically scattered waves. Only this difference, but not the absolute values of the shifts, affects the total three-magnon efficiency, as it is clear from Eqs.~(\ref{e:Vc-tot}) and~(\ref{e:Vs-tot}).   

The last step is the calculation of scattered wave intensity accounting for the damping. Due to the effect of damping, the resulting fused wave is not a monochromatic in $k$-space and its spectrum contains a range of $k_y$ wave numbers, defined by the resonance curve and dispersion relation. Using the approximation of constant primary waves amplitudes, which is valid if the amplitude of the inelastically scattered wave is weaker than that of the primary ones and, thus, has a negligible back-action on them, we consider Hamiltonian equations, Eq.~(\ref{e:dc-dH}), only for the scattered waves. For the confluence it reads
	\begin{equation}
		\frac{dc_\vec k}{dt} + i\omega_\vec k c_\vec k + \Gamma_\vec k c_\vec k = -\frac{2i}{A_\vec k} V_\t{conf} C_\t{i} C_\t{e} e^{-i(\omega_\t{i} + \omega_\t{e})t} \,,
	\end{equation}
where $\Gamma_\vec k$ is the damping rate, $C_\nu$ are envelope amplitudes (i.e., $c_\nu(t) = C_\nu(t) e^{-i\omega_\nu t}$) and index $\vec{k}$ denotes parameters of the particular wave from the spectrum. Stationary solution is $c_\vec k (t) = C_\vec k e^{-i(\omega_\t{i} + \omega_\t{e})t}$ with the envelope amplitude  
	\begin{equation}
		C_\vec k = - \frac{iV_\t{conf} C_\t{i} C_\t{e}}{\e_\vec k \left(i(\omega_k-(\omega_\t{i}+\omega_\t{e}))+\Gamma_\vec k \right)},
	\end{equation}
where we account the norm in Eq.~(\ref{e:norm}), which for the bulk SW is $A_\vec k = 2\e_\vec k$. The output SW is therefore a sum of SWs with different $k_y$. Changing the sum $\sum_{k_y}$ into the integral $L_y/(2\pi)\int dk_y$ and using the linear SW spectrum approximation in the vicinity of the exact three-magnon resonance
	\begin{equation}
		\omega_\vec k= (\omega_\t{i} + \omega_\t{e}) + v_y \Delta k_y ,
	\end{equation}
where $v_y =d\omega_\vec {k}/dk_y$ is the $y$ component of the group velocity and $\Delta k_y$ is the difference between the $y$ component of the wave vector of our scattered waves and $k_{y}$ in case of resonance process with bulk waves. The output SW envelope amplitude is derived as
	\begin{equation}
		C_\t{s}(y) = -iC_\t{i}C_\t{e} \frac{L_y}{2\pi} \int_{-\infty}^{\infty}\frac{V_\t{conf}}{\e_\t{s}} \frac{e^{i\Delta k_y y}}{iv_y\Delta k_y+\Gamma_\vec k} d\Delta k_y .
	\end{equation}
Using the assumption of a weak variation of three-magnon coefficient and ellipticity-related coefficient $\e_\vec k$ within the resonance linewidth, which is natural for low-damping materials, the integration is performed analytically and yields 
	\begin{equation}\label{e:cf-y}
		C_{s}(y) = - \frac{i \tilde V_\t{conf} C_\t{i} C_\t{e}}{\e_\t{s} v_{\t{s},y}} \exp\left[ -\frac{\Gamma_\t{s} y} {v_{\t{s},y}} \right] .
	\end{equation}
Here, we use notation from Eq.~(\ref{e:V123-fin}), i.e., $\tilde V_\t{conf} = V_\t{conf} L_y$. The result does not depend on the system size $L_y$ (see Eq.~(\ref{e:V123-fin})), as it should. Equation (\ref{e:cf-y}) describes bulk SW exponentially decaying from the film edge, generated by the three-magnon scattering at the edge. There is no $x$-dependence as the theory implies infinite uniform plane wave. 

Applying the same procedure, envelope amplitude of an SW, which appears as a result of stimulated three-magnon splitting, is found to be
	\begin{equation}\label{e:cs-y}
	C_{s}(y) = \frac{i \tilde V_\t{split}^* C_\t{i} C_\t{e}^*}{\e_\t{s} v_{\t{s},y}} \exp\left[ -\frac{\Gamma_\t{s} y} {v_{\t{s},y}} \right].
	\end{equation}	 

When analyzing Eqs.~(\ref{e:cf-y})-(\ref{e:cs-y}), we should point two important features. Firstly, damping only affects the decay of the scattered wave as it propagates away from the edge, but it does not affect the amplitude near the edge. This justifies simulations with a reduced damping, which will be presented in the next section. Secondly, it is not only the total three-magnon efficiency (which is sensitive to the phase accumulation during the reflection, Eq.~(\ref{e:Vc-tot})) that affects the amplitude of the scattered wave, but also perpendicular to the edge component of the group velocity of the scattered wave. This is a known effect of radiative losses, which are proportional to the group velocity -- the faster the scattered wave, the greater radiative losses, what leads to the lower wave amplitude. 

Finally, for quantitative comparison of theoretical results with micromagnetic data, we need to convert complex SW envelope amplitudes into real magnetization components. This is done directly from the definition of envelope amplitudes 
	\begin{equation}
		\vec m (y,t) = C_j(y) \vec s_j(\vec r) e^{ -i \omega t} + \t{c.c.} \,,
	\end{equation}
with SW profile $\vec s_j$ defined by Eq.~(\ref{e:s-bulk}) for bulk SWs and (\ref{e:s-edge}) for the edge mode (in the latter case $C_\t{e} \notin f(y)$). 
Then, the amplitude of the  out-of-plane magnetization component of the scattered wave is 
	\begin{equation}\label{e:mz_3}
		m_{s,z} = \frac{|\tilde V| m_{\t{i},z} m_{\t{e}, z}} {2 \e_\t{i} \e_\t{e} v_{\t{s},y}} \exp \left[-\frac{\Gamma_\t{s} y}{v_{\t{s},y}}\right] ,
	\end{equation}
where amplitudes of incident and edge SWs, $m_{\t{i},z}$, and $m_{\t{e},z}$, are defined just before the interaction area. This equation holds for both confluence and stimulated splitting processes with appropriate choice of three-magnon coefficient, $|\tilde V|$, given by Eqs.~(\ref{e:Vc-tot}) or Eq.~(\ref{e:Vs-tot}), respectively.

\section{Results and comparison with simulations}\label{s:res}

We begin our analysis of the results arising from the developed theory with the effect of the incidence angle on the three magnon scattering, which exhibits a nontrivial dependence~\cite{Pawel2022}. We set the frequency of the incident bulk SW frequency to $f_\t{i} = \omega_\t{i}/(2\pi) = 45$~GHz, the frequency of the edge SW frequency to $f_\t{e} = 12$~GHz, thus the waves generated in the confluence and stimulated splitting processes are at frequencies of 57 and 33 GHz, respectively. We vary the incidence angle of the bulk wave, $\varphi$ in the range from 0 to 55$^\circ$ (see the related range of the $k_x$ component of the bulk SW marked in the dispersion relation in Fig.~\ref{fig:geo}c). The inverse localization length $\kappa=62$ \textmu m$^{-1}$ and the precession ellipticity $\e_2=0.55$ for the edge mode are taken, in order to approximate the profile of the edge SW obtained in micromagnetic simulations.

\begin{figure*}
	\centering
	\includegraphics[width=\textwidth]{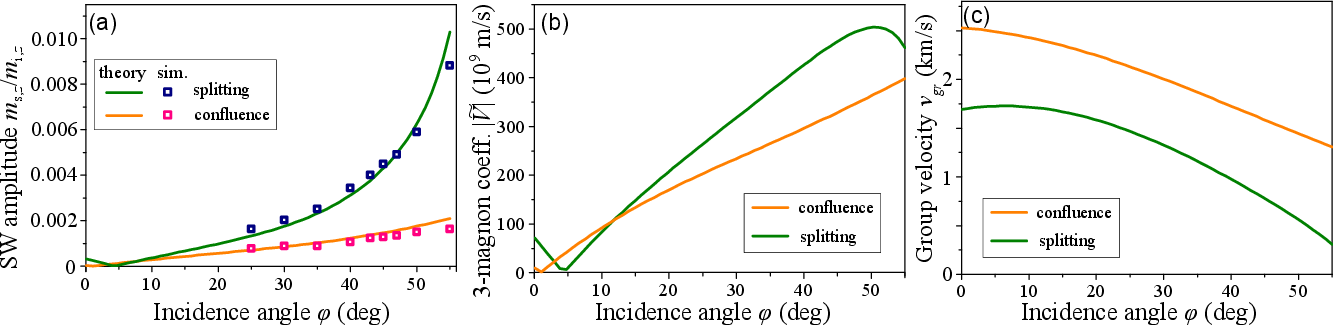}
	\caption{(a) Incidence angle dependence of the scattered wave amplitudes for the confluence and stimulated splitting processes; symbols -- micromagnetic simulations, lines -- theory. (b) and (c) Analytically calculated angular dependencies of the effective three-magnon interaction coefficients based on Eqs.~(\ref{e:Vc-tot}) and~(\ref{e:Vs-tot}) and the group velocity of the scattered SWs. We assumed incident bulk SWs at frequency $f_\t{i}=45$ GHz and the edge SW at $f_\t{e}=12$ GHz.}
	\label{f:mz-phi}
\end{figure*}

The results of micromagnetic results (see the details of simulations in the Appendix), showing the amplitudes of inelastically  scattered waves at the distance of $y = 100$~nm from the film edge, are presented in Fig.~\ref{f:mz-phi}(a) as symbols. For the both three-magnon processes -- confluence and stimulated splitting -- the scattered wave amplitude increases with an increase of  $\vp$. However, for an angle $\vp > 57^\circ$, the energy and momentum conservation rules for the splitting process cannot be satisfied (see the dispersion relation in Fig.~\ref{fig:geo}c).  Thus, above this cut-off angle, only the confluence and nonresonant stimulated splitting processes are possible, and the efficiency of the latter is much smaller than that of the resonant processes. We observe a significant quantitative difference in the scattered waves amplitudes. The wave resulting from stimulated splitting has a larger amplitude than that caused by the confluence, and this difference becomes more pronounced at larger incidence angles, reaching the ratio $m_{s,z}^{\text{split}}/m_{s,z}^{\text{conf}}=3.4$ at $\vp=50^\circ$.  Analytical calculations according to Eq.~(\ref{e:mz_3}), shown by the solid lines, demonstrate a very good agreement with the simulation results, enabling us to deeply inspect three-magnon processes and their asymmetry.

According to Eq.~(\ref{e:mz_3}), the principal characteristics that vary with the incidence angle and affect the scattered wave amplitude (normalized by the incident wave amplitudes), are the three-magnon interaction efficiency $|\tilde V|$, and the projection of the group velocity of the scattered wave on the $y$-axis, $v_{s,y}$. The variation of two other parameters -- the ellipticity-related coefficient of the incident bulk wave, $\e_\t{i}$, and the damping rate of the scattered wave, $\Gamma_\t{s}$ -- with the angle $\vp$, is marginal. Angular dependence of the effective three-magnon coefficients $|\tilde V|$, which account for two elementary three-magnon scattering processes as described in Eqs.~(\ref{e:Vc-tot}) and (\ref{e:Vs-tot}), is shown in Fig.~\ref{f:mz-phi}(b). For both confluence and stimulated splitting, $|\tilde V|$ is a monotonic increasing function of $\vp$, leading to an increase in the scattered SW amplitudes with $\vp$ for the main part of the studied angular range. This nearly monotonic dependence stems from the fact, that the three-magnon interaction is determined by the off-diagonal component of the dipolar tensor $N_{\vec k}^{yx}$ [Eq.~(\ref{eq:N_dip})], which is proportional to $ k_x k_y \sim \sin(2\vp)$. Thus, one might expect an increase in $|\tilde V|$  in the range $\vp \in [0,45^\circ]$, followed by a decrease, as was also shown by an approximate qualitative model in Ref.~\cite{Pawel2022}. Of course, the interplay of the different terms in Eq.~(\ref{e:V123-fin}) and the interference effect lead to more complicated dependencies, as we observe in Fig.~\ref{f:mz-phi}(b). 
 In particular, we observe the interesting phenomenon of the three-magnon interaction vanishing for the stimulated splitting at an angle $\vp \approx 5^\circ$. The explanation for this phenomenon is provided below. 

At the same time, in the range $ 20^\circ <\vp < 55^\circ$,  $|\tilde V|$ is  similar for both processes (the ratio $|\tilde V_\text{split}|/|\tilde V_\text{conf}|$ is approximately 1.4 at  $\varphi=50^\circ$). Therefore, it cannot  result in the significantly different scattered SW amplitudes of the stimulated splitting and confluence processes. The scattered wave group velocities, i.e., their projections $v_{s,y}$, differ considerably over the whole range of incidence angles (see Fig.~\ref{f:mz-phi}(c)). This is especially apparent close to the cut-off angle $\vp \approx 57^\circ$, i.e., $v_{s,y}^\text{split}/v_{s,y}^\text{confl}=0.33$  at  $\varphi=50^\circ$. This difference is the dominant reason for the more intensive stimulated splitting process compared to the confluence one.

\begin{figure}
	\includegraphics[width=0.9\columnwidth]{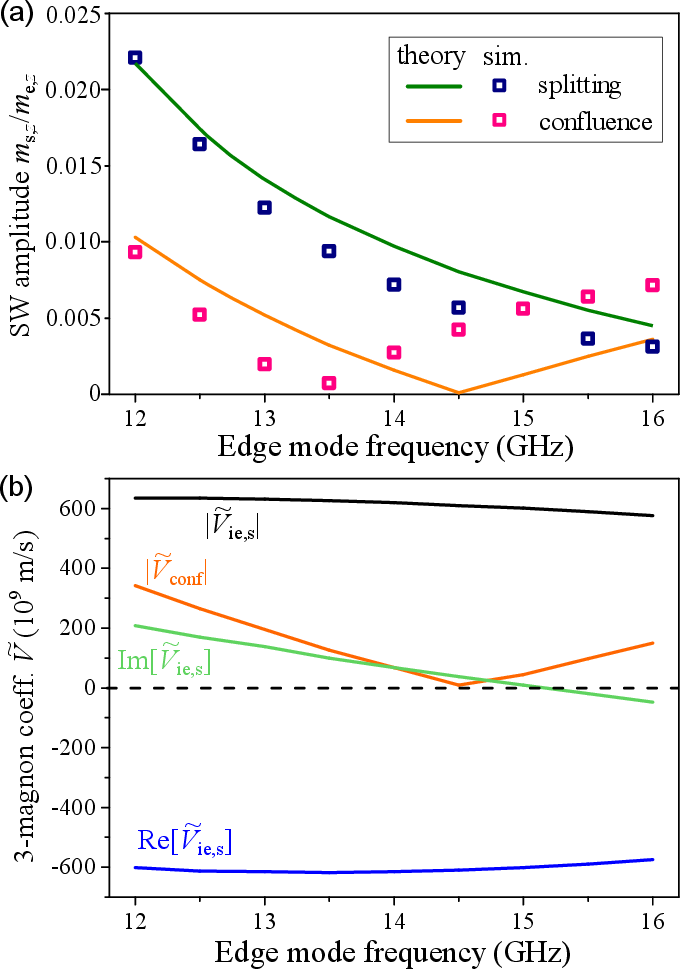}
	\caption{(a) Normalized amplitude of the inelastically scattered SW in dependence on the edge SW frequency: symbols --  micromagnetic simulations, lines --  theory. (b) Edge SW frequency dependencies of the total effective three-magnon coefficient for the confluence process $|\tilde V_\t{conf}|$ (orange line), modulus of the elementary confluence process efficiency $|\tilde V_\t{ie,s}|$ (black line), and real (blue) and imaginary (green) parts of elementary confluence process efficiency. Incident bulk SW frequency is $f_\t{i} = 45$~GHz, and the angle of incidence is $\vp = 30^\circ$.} 
	\label{f:mz-fe}
\end{figure}

Next, we consider the effect of the edge mode frequency on the three-magnon scattering. The incidence angle of the bulk SW is fixed to $\vp = 30^\circ$, its amplitude was $m_{\t{i},z} = 0.039$. We varied the frequency of the edge mode within the range $12-16$~GHz (see Fig.~\ref{f:mz-phi}(c)).  For these frequencies, the inverse localization length $\kappa$ varies from $62$ to $72\ \mu\text{m}^{-1}$, while the precession ellipticity $\varepsilon_2$ ranges from $0.55$ to $0.68$. Although the excitation field was fixed, the edge mode amplitude changes with the frequency, from $m_{\t{e},z} = 0.04$ to $m_{\t{e},z} = 0.06$. We therefore present the ratio of the scattered-to-edge SW amplitudes in Fig.~\ref{f:mz-fe}(a). The simulation results, shown as empty dots, evidence a strong influence of the edge mode frequency on the amplitude of the scattered waves. This is especially pronounced for the confluence wave, where the scattered mode amplitude almost vanishes at $f_\t e =13.5\,$~GHz. This behavior is non-trivial, as there are no peculiarities in the group velocity of the scattered wave in the studied frequency range. Furthermore, it is difficult to imagine that the three-magnon process is prohibited for a particular SW frequency (at least, expression (\ref{e:V123-fin}) does not provide a clear explanation for this possibility).

Theoretically calculated scattered mode amplitudes, shown in  Fig.~\ref{f:mz-fe}(a) with solid lines, reasonably agree with the micromagnetic data. In particular, the calculations also give vanishing fused wave amplitude, although at a somewhat higher frequency of the edge SW, i.e., $f_\t{e} \approx 14.5$~GHz (the reasons of discrepancies will become clear below). To deeply analyze peculiarities of three-magnon interaction, we plot effective three-magnon coefficient $|\tilde V_\t{conf}(f_\text{e})|$ in Fig.~\ref{f:mz-fe}(b) (orange line). Indeed, the coefficient goes to zero at $14.5$~GHz, clearly showing its responsibility for the vanishing amplitude of the fused wave. According to Eq.~(\ref{e:Vc-tot}), this coefficient depends on the  two elementary three-magnon processes (shown schematically in Fig.~\ref{fig:2}). If looking on the frequency dependence of the efficiency of one elementary confluence process, for instance $|\tilde V_\t{ie,s}|$, one finds no peculiarities, only a weak smooth decrease with $f_\t{e}$ (black line in Fig.~\ref{f:mz-fe}(b)). As we have pointed out above, such dependence is quite expected -- the wide spatial spectrum of the edge mode allows three-magnon interaction with bulk waves with almost no restrictions in the general case.

The reason of vanishing effective three-magnon coefficient is, thus, the competition of two terms in Eq.~(\ref{e:Vc-tot}). It can be proven that the reversal of $k_y$ components of both the bulk SWs, involved in a three-magnon process, does not change the modulus of three-magnon coefficient, but affects its argument. Namely, in the used notations, $V_{\bar{\t{i}}\t{e},\bar{\t{s}}} = - V_\t{ie,s}^*$, i.e., these coefficients have the same imaginary parts but opposite real parts. This means scattered wave, which appears via the interaction of incident and edge waves (see the process shown in Fig.~\ref{fig:2}(b)), has different phase respectively to the wave, which is caused by three-magnon scattering after reflection from the edge (see, Fig.~\ref{fig:2}(a)). The interference of scattered SWs coming from two elementary three-magnon processes is determined by this phase shift, resulting in a complex behavior of the total scattering efficiency.

In an ideal scenario with a sharp boundary, the phase shift acquired by a bulk SW reflection is the same for all the waves and is equal to $\pi$ radians. Therefore, the effective three-magnon coefficient is $V_\t{conf}  = - (V_{\bar{\t{i}}\t{e},\bar{\t{s}}} + V_\t{ie,s}) = -2\t{Im}[V_\t{ie,s}]$. Thus, in this idealized case, only the imaginary part of the three-magnon coefficient $V_\t{ie,s}$ matters, and it goes to zero at $f_\t{e} \approx 15.5\,$~GHz (see the green line in Fig.~\ref{f:mz-fe}(b)). In reality, an arbitrary phase accumulation, acquired by the incident and scattered SWs upon the reflection, affects an SW interference and shifts $|\tilde V_\t{conf}|=0$ point to a lower frequency of the edge mode. It is the same physics of SW interference that causes the split wave to vanish at $\vp \approx 5^\circ$ in Fig.~\ref{f:mz-phi}(a). Similarly, as well as the pronounced dependence of the split mode amplitude on $f_\t{e}$ observed in Fig.~\ref{f:mz-fe}(a).

The sensitivity of the total three-magnon process efficiency to SW phase accumulation can only be estimated using analytical methods. This is mainly due to the non-local nature of dipolar interactions, as well as the extended area over which elementary scattering processes take place. At the same time, the phase sensitivity of inelastically scattered SWs opens up the possibility of controlling the scattering of three magnons of bulk and edge SWs. For example, applying a local magnetic field near the film edge affects the dispersion and localisation of the edge mode, as well as the phase accumulation upon bulk wave reflection. Three-magnon scattering may be more sensitive to this effect. Additionally, modifying the magnetic anisotropy at the film edge, for example, by applying mechanical strain, could be another approach, although this might be challenging for thin films.

\section{Summary}

In summary, this work provides a comprehensive analysis of the inelastic scattering of bulk and edge SWs at the edge of an in-plain magnetized thin ferromagnetic film,  including both the three-magnon confluence and stimulated splitting processes. The localized nature of the edge mode, and consequently its wide spatial spectrum, results in a breaking of the momentum conservation law in the perpendicular to the edge direction. Consequently, three-magnon interaction is possible over a broad frequency range and for a wide range of incidence angles, in contrast to the nonlinear scattering of three bulk SWs.

The developed theory of the three-magnon scattering of bulk and edge SWs, verified by micromagnetic simulations, allows for the identification of the main three factors that determine the amplitude of inelastiaclly scattered waves:
\begin{itemize}
	\item The efficiency of an elementary scattering process, $V_\t{ie,s}$. It is determined by the off-diagonal component of the dipolar interaction tensor and follows the trend of $\sim \sin[2\vp]$. In the general case, however, it is a more complex function of the incident angle as well as the frequencies and wave vectors of the SWs. 
	\item The group velocity of the scattered wave  $v_{\t{s},y}$, that is, its component normal to the edge of the film. The amplitude of the inelastic scattered wave is inversely proportional to $v_{\t{s},y}$. This factor is the main reason for the significant difference in the amplitudes of waves generated in the stimulated splitting and confluence processes.  In other words, the amplitude of the wave generated in the splitting process is larger, and the angular dependencies of the scattered wave amplitude are more complex than those given by $V_{\t{ie},s}$.
	\item The interference of two partial waves (or four in the general case) generated by the incident SW in three-magnon scattering processes: (i) before the reflection and (ii) after the elastic reflection from the edge of the film. This interference is sensitive to the phase condition upon reflection, which is frequency-dependent due to the presence of the field well. This interference gives rise to unexpected angular and frequency dependencies in the amplitude of the inelastically scattered SW, which could eventually result in scattering vanishing.
\end{itemize}	

Our results show that the nonlinear interaction between the bulk and edge or other localised SWs exhibits rich physics and is sensitive to the conditions at the edge of the film and, consequently, to external influences. Thus, it can potentially be controlled and be useful for various magnonic applications.

\begin{acknowledgements}
This work was supported by Polish National Science Centre projects 2020/37/B/ST3/03936, 2022/45/N/ST3/01844 and 2021/43/I/ST3/00550, and by the National Academy of Sciences of Ukraine, Grant No. 08/01-2024(5). The numerical simulations were performed at the Poznan Supercomputing and Networking Center (Grant No. 398).
\end{acknowledgements}

\appendix*
\section{Micromagnetic simulations}
To validate our theoretical findings, we employ micromagnetic simulations using MuMax3 software~\cite{Vansteenkiste_AIPAdv2014}. It solves the Landau-Lifshitz-Gilbert equation with finite difference time domain method. In contrast to spontaneous splitting, which possesses a damping-defined threshold, the damping does not directly affect three-magnon confluence and stimulated splitting, as shown in Eq.~(\ref{e:mz_3}). Thus, in order to simplify the numerical analysis, we set a reduced Gilbert damping parameter of $\alpha_G=0.0001$. Lower damping allows for a longer propagation of scattered SW beams from the edge, where the beams become more clear spatially separated and, thus, easier for accurate numerical analysis. 

In the simulations, both the edge SW mode and bulk SWs are excited by localized rf magnetic fields in the form of the beams. These fields were introduced as harmonically oscillating, analytical functions representing two independent antennas within the system.

The first antenna, positioned in the bulk of the layer, comprised two harmonic magnetic fields phase-shifted in such a way as to achieve constructive interference in only one direction. Following the approach presented in~\cite{whitehead2019graded}, such a field is defined as
\begin{equation}
    \begin{split}
        B_{\mathrm{ext}, y}(t,x',y') = A(1-e^{-0.002 \pi f_\t{i} t})G(y')G(x') \\
        \times [  \mathrm{sin}(k_x x')\mathrm{sin}(2 \pi f_\t{i} t) + \mathrm{cos}(k_x x')\mathrm{cos}(2 \pi f_\t{i} t) ],
    \end{split}
    \label{eq:sw}
\end{equation}
where $A=0.01B_0$ is the amplitude of the dynamic field, $G(\zeta')=\mathrm{exp}(-\frac{\zeta'^2}{4\sigma_\zeta^{2}})$ is a Gaussian function defining antenna's shape along the antenna axes with width of $\sigma_\zeta$, $\zeta$ stands for either $x$ or $y$ coordinate.
The bulk SWs were modulated by a Gaussian envelope with a full width at half maximum (FWHM) of 760~nm, allowing them to be treated as beams. In the simulated weakly dispersive system, these beams remained well-collimated throughout the simulation. The propagation angle of the bulk SWs was controlled by rotating their antenna with respect to the system's interface. 
The edge SWs were excited by the second antenna, modeled as a point source with a two-dimensional Gaussian envelope of FWHM 35~nm.


Then the simulations were run until the system reached the steady-state. The simulation time were changed depending on the angle of incidence, the beams with high angle of incidence propagated along longer trajectories what extended the simulation time. In the performed simulations the times ranged from $240$~ns to $300$~ns for angles of incidence $25^{\circ}$ and $60^{\circ}$ respectively. After reaching the steady-state $800$ magnetization configurations were saved with $5$~ps time-steps, what yield frequency resolution of $250$~MHz. Such a resolution allows for spectral analysis of inelastically scattered SWs considered in this paper as their frequencies are multiples of $250$~MHz.


%

\end{document}